\documentstyle[epsfig]{aipproc}
\begin{document}
\vspace*{3cm}
\title{Parameterization of Proton-Proton Total Cross Sections from 10 GeV to 100 TeV } 
\author{J. P\'{e}rez-Peraza$^{\dagger}$,
J. Velasco$^*$,
A. Gallegos-Cruz$^{**}$\\
M. Alvarez-Madrigal$^{\dagger}$, 
A. Faus-Golfe$^*$, and
A. S\'{a}nchez-Hertz$^{\dagger}$}
\address{$^{\dagger}$Instituto de Geofisica, UNAM, 04510, C.U., 
Coyoacan, Mexico D.F.. MEXICO\\
$^*$IFIC, Centro Mixto CSIC-Universitat de Valencia, Doctor Moliner 50,\\
46100 Burjassot, Valencia, SPAIN\\ 
$^{**}$Ciencias Basicas, UPIICSA, I.P.N., Te 950, 
Iztacalco,08400, Mexico D.F, MEXICO}
\maketitle
\begin{abstract} 
Present estimations of proton-proton total cross sections at very high energies are obtained 
from cosmic rays  ($ > 10^{17}$ eV) by means of some approximations
and  the knowledge of the measured proton-air cross
section at these energies. Besides, total cross sections are measured with present day high 
energy colliders up to nearly $2$ TeV in the center of mass ($\sim\ 10^{15}$ eV in the 
laboratory). 
Here we use a phenomenological model based on the Multiple-Diffraction  approach
to succesfully describe data at accelerator energies. Then we
 estimate with it proton-proton total cross sections at cosmic ray energies.
On the basis of a 
forecasting regression analysis we determine confident error bands, analyzing 
the sensitivity of our predictions  to the employed data for extrapolation. 
\end{abstract}
\newpage
\setcounter{page}{1}
\section*{Introduction}
Recently a number of difficulties 
in uniting accelerator and cosmic ray 
values of hadronic cross-sections within the frame of the highest up-to-date data
have been summarized \cite{Engel1}. 
Such united picture appears 
to be highly important for at least, the interpretation of results of 
new cosmic ray experiments, 
as the HiRes \cite{hires} and in designing proposals that are currently in progress, 
as the Auger 
Observatory \cite{Auger}, as well as in designing detectors for future accelerators, 
as the CERN pp 
Large Hadron Collider (LHC). Although most of accelerator measurements 
of $\sigma_{tot}^{\bar{p}p}$ and $\sigma_{tot}^{pp}$ at 
center of mass energy $\sqrt{s}$  $\leq$ 1.8 TeV are quite 
consistent among them, this is 
unfortunately not the case for cosmic ray experiments at $\sqrt{s} > 6 $ TeV 
where some 
disagreements exist among different experiments. This is also the case 
among different predictions 
from the extrapolation of accelerator data up to cosmic ray energies: 
whereas some works predict 
smaller values of $\sigma_{tot}^{pp}$ than those of cosmic ray experiments 
(e.g. \cite{DL1,Augier1}) other 
predictions  agree at some specific energies with cosmic ray results 
(e.g. \cite{Block1,MM1}). Dispersion 
of cosmic ray results are mainly associated to the strong model-dependence 
of the relation between 
the basic hadron-hadron cross-section and the hadronic cross-section in air.  
The later determines 
the attenuation  lenght of hadrons in the atmosphere, 
which is usually measured in different ways, 
and depends strongly  on the rate ($k$) of energy dissipation 
of the primary proton into the 
electromagnetic shower observed in the experiment: such a cascade 
is simulated by different Monte 
Carlo techniques implying additional discrepancies between different experiments. 
Furthermore, 
$\sigma_{tot}^{pp}$ in cosmic ray experiments is determined 
from $\sigma_{p-air}^{inel}$ using a 
nucleon-nucleon scattering amplitude which is frequently in disagreement 
with most of accelerator 
data \cite{Engel1}. On the other hand, we dispose of parameterizations 
(purely theoretically, empirical or 
semi-empirical based) that fit pretty well the accelerator  data. 
Most of them agree that at the 
energy of the LHC (14 TeV in the center of mass) or higher (extrapolations) 
the rise in energy of 
$\sigma_{tot}^{pp}$ will continue, though the predicted values differ 
from model to model. We 
claim that both the cosmic ray and parameterization approaches must complement 
each other in order 
to draw the best description of the hadronic cross-section behavior 
at ultra high energies. 
However, the present status is that due to the fact that interpolation 
of accelerator data is 
nicely obtained with most of parameterization models, it is expected 
that their extrapolation to 
higher energies be highly confident: as a matter of fact, 
parameterizations are usually based in a 
short number of fundamental  parameters, in contrast with 
the difficulties found in deriving 
$\sigma_{tot}^{pp}$ from cosmic ray results \cite{Engel1}. 
If extrapolation from parameterization models
is correct this would imply that $\sigma_{p-air}^{inel}$ should be smaller, 
which would have 
important consequences for development of high energy cascades. 
\par
With the aim of contributing to the understanding of this problem,
in this paper we first briefly analyze in the first two sections 
the way estimations are done  
for  proton-proton total cross sections
from  accelerators as well as from cosmic rays.  
We find serious discrepancies among both estimations.
In the third section, on the basis 
of the Multiple Diffraction model  applied to accelerator data, we predict 
$pp$ total cross section values with smaller errors than with the standard techniques.
We conclude 
with a discussion about the quality of present cosmic ray  estimations.
\section{Hadronic $\sigma_{\lowercase{tot}}^{\lowercase{pp}}$ from accelerators}
Since the first results of the Intersecting Storage Rings(ISR) 
at CERN arrived in the 70s, it is a 
well established fact that $\sigma_{tot}^{pp}$ rises with energy (\cite{Amaldi1,ISR1}). 
The CERN 
$S\bar{p}pS$ Collider  found this rising valid for $\sigma_{tot}^{\bar{p}p}$ 
as  well \cite{UA41}.
 Later, the Tevatron confirmed that for $\sigma_{tot}^{\bar{p}p}$ 
the rising still continues at 
1.8 TeV, even if there is a disagreement among the  diferent  experiment values
as for the exact value (\cite{E710,CDF}). A thoroughful discussion 
on these problems may be found in  \cite{GM2,Blois99}.
It remains now 
to estimate the amount of rising of
the total cross section at those energies. 
Let us resume the standard technique used by 
accelerator experimentalists \cite{Augier1}.
\par
Using  a semi-empirical parameterization based on Regge theory  and asymptotic
theorems experimentalists   have succesively described their data from the ISR 
to the $S\bar{p}pS$ energies.  It takes into account  all the available data  for
$\sigma_{tot}^{pp}$, $\rho^{pp}$, 
$\sigma_{tot}^{\bar{p}p}$ and $\rho^{\bar{p}p}$, 
where  $\rho^{pp,\bar{p}p}$,   is 
the real part of the ($pp, \bar{p}p$) forward elastic amplitude at $t=0$. 
The fits are performed using the once-subtracted  dispersion relations:
\begin{equation}
\rho_{\pm}(E)\sigma_{\pm}(E) = {{C_{s}}\over {p}}
+ {{E}\over {\pi p}} \int_{m}^{\infty} dE' p'
\left[{{\sigma_{\pm}(E')}\over {E'(E'-E)}} -
 {{\sigma_{\mp}(E')}\over {E'(E'+E)}} \right]
\end{equation}
where $C_{s}$ is the substraction constant. 
The expression for $\sigma_{tot}^{pp,\bar{p}p}$ is: 
\begin{equation}
\sigma_{-,+}^{tot} = A_{1}E^{-N_{1}} \pm A_{2}E^{-N_{2}} + C_{0} +
                    C_{2}[ln(s/s_{0})]^{\gamma}
\end{equation}                  
where - (+) stands for $pp$ ($\bar{p}p$)  scattering.  
Cross sections are measured in mb and energy in GeV, E being the energy
measured in the lab frame. The scale factor $s_{0}$ have been arbitrarily
chosen equal to 1 GeV$^{2}$.  
The most interesting piece
is the one controling the high-energy behaviour, 
given by a $ln^{2}(s)$ term, in
order to be compatible, asymptotically, with the Froissart-Martin bound
\cite{FM1}.
The parameterization assumes 
$\sigma_{tot}^{pp}$ and $\sigma_{tot}^{\bar{p}p}$ to be the same
asymptotically. This is justified  from the very precise measurement  of the
$\rho$ parameter  at 546 GeV \cite{Augier2}. 
\par
The eight free parameters are 
determined by a fit 
which minimizes the $\chi^{2}$ function 
\begin{equation}
\chi^{2} = \chi^{2}_{\sigma_{\bar{p}p}}+\chi^{2}_{\rho_{\bar{p}p}}+
         + \chi^{2}_{\sigma_{pp}}+\chi^{2}_{\rho_{pp}}
\end{equation}
The fit has proved its validity 
predicting from the ISR data (ranging from 23  to 63 GeV in the center of mass), 
the $\sigma_{tot}^{\bar{p}p}$ 
value  later found at the $S\bar{p}pS$ Collider, 
one order of magnitude higher in energy (546 GeV)
\cite{Amaldi2,UA41}. With the same well-known technique 
and using the most recent results it is possible 
to get estimations for $\sigma_{tot}^{pp}$ at the  
LHC and higher  energies. These 
estimations, together with our present experimental knowledge 
for both $\sigma_{tot}^{pp}$ and 
$\sigma_{tot}^{\bar{p}p}$ are plotted in figure 1. 
We have also plotted 
the cosmic ray experimental data from AKENO (now AGASSA) \cite{Akeno1} 
and the Fly's Eye experiment 
\cite{FE1,FE2}. The curve is the result of the fit described in \cite{Augier1}. 
The increase in 
$\sigma_{tot}^{pp}$ as the energy increases is clearly seen. 
Numerical predictions  from this analysis are given  in Table 1.
It should be remarked that 
at the LHC energies and beyond the fitting results display relatively 
high error values, equal or bigger than 8 mb. 
We conclude  that it is necessary to look for ways to reduce the  errors 
and make the extrapolations more precise.
\begin{center}
\begin{figure}[b!]
\centerline{\epsfig{file=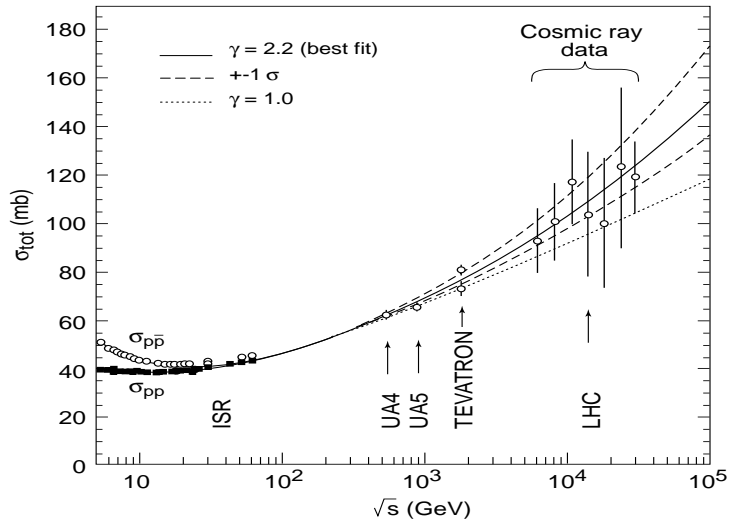,height=2.7in,width=3.8in}}
\caption{Experimental $\sigma_{tot}^{pp}$ and $\sigma_{tot}^{\bar{p}p}$ with the
 prediction of [5].}
\end{figure}
\end{center}
\begin{center}
\begin{table}
\caption{$\sigma_{tot}^{\bar{p}p}$ data from high energy accelerators: fits values are from [5].}
\begin{tabular}{|l||l||r|} \hline
$\sqrt s$ (TeV)& &$ \sigma_{tot}^{pp}$ (mb) \\ \hline 
0.55 & Fit  &  $61.8  \pm 0.7 $\\
     & UA4  &  $62.2  \pm 1.5 $\\
     & CDF  &  $61.5  \pm 1.0 $\\ \hline             
1.8  & Fit  &  $76.5  \pm 2.3 $\\
     & E710 &  $72.8  \pm 3.1 $\\
     & CDF  &  $80.6  \pm 2.3 $\\  \hline
14   & Fit  & $109.0  \pm 8.0 $\\ \hline
30   & Fit  & $126.0  \pm 11.0 $\\ \hline
40   & Fit  & $130.0 \pm 13.0 $\\ \hline 
\end{tabular}
\end{table}
\end{center}
\section{Hadronic $\sigma_{\lowercase{tot}}^{\lowercase{pp}}$ from cosmic rays}
 Cosmic rays experiments give us  $\sigma_{tot}^{pp}$  in an indirect way:
we  have to derive it from cosmic ray extensive air 
shower (EAS) data. But, as summarized in \cite{Engel1} 
and widely discussed in the literature, the 
determination of $\sigma_{tot}^{pp}$ is a rather complicated 
process with at least two well 
differentiated steps. In the first  place,  the primary interaction 
involved in EAS is  proton-air; what it is 
determined through EAS is the $p$-inelastic cross section, $\sigma_{inel}^{p-air}$, 
through some 
measure of the attenuation of the rate of showers, $\Lambda_{m}$,  deep in the atmosphere:
\begin{equation}
\Lambda_{m} = k \lambda_{p-air}  = k \ \frac {14.5 m_{p}}  {\sigma_{inel}^{p-air}}
\end{equation}
The  $k$ factor  parameterizes the rate at which the energy  
of the primary proton is dissipated 
into electromagnetic energy. A simulation with a full representation 
of the hadronic interactions 
in the cascade is needed to calculate it. This is done 
by means of Monte Carlo techniques \cite{Pryke,Hillas,Fletcher}.
Secondly, the connection between $\sigma_{inel}^{p-air}$ and $\sigma_{tot}^{pp}$
is model dependent. A theory for nuclei interactions must be used. 
Usually is Glauber's theory \cite{G1,GM1}.
The whole procedure makes hard to get a general agreed value for $\sigma_{tot}^{pp}$. 
Depending on the 
particular assumptions made the values may oscillate by  large amounts, from as low to 
$133 \pm 10$ mb \cite{Akeno1} to nearly $165 \pm 5$ mb \cite{Niko}
 and even $175_{-27}^{+40}$ \cite{GSY} at 
$\sqrt s = 40 $ TeV.  
\par
>From this analysis the conclusion is 
that cosmic-ray estimations of 
$\sigma_{tot}^{pp}$ are not of much help to constrain extrapolations 
from accelerator energies  \cite{Engel1}. Conversely we could ask if those extrapolations 
could not be used to constrain  cosmic-rays estimations.
\section{A Multiple-Diffraction Approach For 
$\sigma_{\lowercase{tot}}^{\lowercase{pp}}$ } 
Let us tackle the 
mismatching between accelerator and cosmic ray estimations using 
the multiple-diffraction model \cite{GV1}. 
The elastic hadronic scattering amplitude for the collision of two hadrons A and B
is described  as  
\begin{equation}
F(q,s) = i \int_{0}^{\infty } b \ db \left[ 1 - e^{i \Xi (b,s)}
 \right]  J_{o}(qb) 
\end{equation}
where $\Xi (b,s)$ is the eikonal, $b$ the impact parameter, $J_{o}$
the zero-order Bessel function and $q^{2} = - t$ the four-momentum transfer squared. 
The eikonal can be expressed at first order
as $\Xi (b,s) = \left< G_{A}G_{B} f \right> $, where $G_{A}$ and
$G_{B}$ are the hadronic form factors, $f$ the averaged elementary
amplitude among the constituent partons and the brackets denote 
the symmetrical two-dimensional Fourier transform. 
Given the elastic amplitude $F(q,s)$, 
$\sigma_{tot}^{pp}$ may be evaluated with the help of the optical theorem: 
\begin{equation}
\sigma_{tot}^{pp} = 4 \ \pi \ ImF(q=0, s)
\end{equation}
Multiple-diffraction models differ one from another by the particular choice of 
parameterizations made for $G_{A}$ and $G_{B}$ and the elementary 
amplitude $f$. In the case of identical particles, as is our case, $G_{A}=G_{B}=G$.
For our purposes we follow the parameterization developed in \cite{MM1}
which has the advantage of using a small set of free parameters, five in total:
two of them $(\alpha^{2}, \beta^{2})$ associated with the form factor $G$
\begin{equation}
   G = (1+ \frac{q^{2}}{\alpha^{2}})^{-1} (1+ \frac{q^{2}}{\beta^{2}})^{-1} 
\end{equation}
and three energy-dependent parameters $(C, a, \lambda )$ associated with the 
elementary complex amplitude $f$  
\begin{equation}
 f(q,s) = Ref(q,s)+ iImf(q,s) 
\end{equation}
\begin{equation}
 Imf(q,s)= C \frac{1-\frac{q^{2}}{a^{2}}}{1-\frac{q^{4}}{a^{4}}}  \ ; 
 Ref(q,s) = \lambda(s)Imf(q,s)
\end{equation}
We get 
\begin{equation}
ImF(q=0,s) \ = \int_{0}^{\infty} \left[ 1 - e^{- \Omega(b,s)}
 \cos \left\{ \lambda \Omega(b,s) \right\} \right] b \ db \ J_{o}(q,b)
\end{equation}
with the opacity $\Omega(b,s)$  given as:
\begin{equation}
\Omega(b,s) = \int_{o}^{\infty} G^{2} \ Imf(q,s)\ J_{o}(q,b) \ qdq 
\end{equation}
\begin{eqnarray}
\Omega(b,s) & = & C \{ E_{1}K_{0}(\alpha b) + E_{2}K_{0}(\beta b) + E_{3}K_{ei}(ab) + \nonumber \\
        & & E_{4}K_{er}(ab) + b \left[ E_{5}K_{1} (\alpha b) + E_{6}K_{1} (\beta b) \right] \ \}
\end{eqnarray}
where   $k_{0}, k_{1}, k_{ei}, $  and $ k_{er}$ are the modified Bessel  functions, 
and $E_{1}$  to $E_{6}$  are functions of the free parameters. 
The proton-proton total cross-section is directly determined by the expression
\begin{equation}
\sigma_{tot}^{pp} = 4\pi \int_{o}^{\infty} b \ db  \left\{ 1 - e^{- \Omega(b,s)}
 \cos \left[ \lambda \Omega(b,s) \right] \right\} \ J_{o}(q,b)
\end{equation}             
This equation  was numerically solved. The overall procedure  is done in a three step process.
\begin{itemize}
\item
First, we determine the parameters of the model by fitting all 
the  
$pp$ as well as  $\bar{p}p$ 
accelerator data
(differential elastic cross sections
and $\rho$ values). 
in the interval $13.8 \leq \sqrt s \leq 1800$ GeV. 
The obtained values
are listed in Table 2.  
\item
Secondly, and most important, 
we estimate an error band  for each of the energy-dependent parameters. 
To this end we introduce  the so-called forecasting technique,
based   on the multiple linear regression theory. It consists 
in  determining a prediction  equation por each free parameter. This allows  
to set  a confidence band  for each parameter and  the  confidence  band
for the predicted total cross section.
The technique is explained in detail elsewhere \cite{PP1}.
\item
Finally, we proceed to extrapolate our results to high energies.
Results  are summarized  in Table 3a  and plotted in figure 2b, 
together with cosmic ray data. 
As a comparison, we list in Table 3b 
the extrapolated values obtained 
when  only $pp$ data, covering a much smaller t-range interval
($13.8 \le \sqrt{s} \le  62.5 $ GeV),  was used.  This was the  method 
in \cite{MM1}, but their extrapolated values were given 
without quoting any errors.
\end{itemize} 
It may be argued that $\sigma_{tot}^{pp}$  and
$\sigma_{tot}^{\bar{p}p}$ are different at high energies:
This is the ``Odderon hypothesis'', which as indicated in Section I,
has been very much weakened  \cite{Augier2}.
Taking this into account, in our multiple-diffraction analysis 
it is assumed the same behaviour
for $\sigma_{tot}^{pp}$ and $\sigma_{tot}^{\bar{p}p}$ at high energy. 
\par
Of course, if we limit our fitting calculations to the 
accelerator domain $\sqrt s \leq 62.5 $ GeV (Table 3b), 
our results are the same as those 
obtained in \cite{MM1}. 
In that case, 
the $\sigma_{tot}^{pp}$ values obtained when extrapolated 
to ultra high energies seem to confirm 
the highest quoted values of the cosmic ray experiments \cite{Niko,GSY}. 
That would imply the 
extrapolation cherished by experimentalists is wrong. 
But the prediction  $\sigma_{tot}^{pp} = 91.6 $ mb 
at the Fermilab Collider energy  
(1.8 TeV) is too high, first, and secondly, difficult to interpret,
as no error is quoted in that work. 
In Table 1 we see that 
the measured $\sigma_{tot}^{\bar{p}p}$ at 546 GeV is  smaller than the
predicted $\sigma_{tot}^{pp}$ by near 8 mb, and in the case of 1.8 TeV
by more than 15 mb, which no available model
is able to explain \cite{Blois99}. 
Also it can be noted that the extrapolation from figure 2a
to ultra high energies may claim agreement 
with the analysis 
carried out in \cite{Niko} and the experimental data 
of the Fly's Eye \cite{GSY}, and even with the 
Akeno collaboration \cite{Akeno1}, because its errors are so big that 
overlap with the errors reported in \cite{GSY}. 
That is, such an extrapolation, Fig.2a, produces an error band so large 
at cosmic ray energies that any
cosmic ray results become compatible with results at accelerator energies, 
as it is claimed in 
\cite{MM1}. However, when additional data at higher accelerator energies 
are included (Table 3b), 
both the predicted values and the error band obviously change. 
This can be clearly 
seen in figure 2b, 
where we have considered data at 0.546 TeV and 1.8 TeV 
(see Table 1) in which case the predicted value of $\sigma_{tot}^{pp}$ 
from our extrapolation at 
$\sqrt s = 40 $ TeV, $\sigma_{tot}^{pp} = 131.7_{-4.6}^{+4.8}$ mb 
is incompatible with 
those in \cite{Niko,GSY} by several standard deviations, 
though no so different to the Akeno results 
and the predicted value in \cite{Augier1}. 
\par
Concerning the quoted  error bands, the forecasting technique  
has reduced the errors, as is seen in figure 2(b), 
nearly by a factor of 3, as compared with the results quoted  in Table 1.  
\begin{center}
\begin{table}
\caption{Values of the  parameters C, $\alpha^{-2}$, $\lambda$ at each energy.
They are obtained from fitting  the $pp$ ($\bar{p}p$) differential cross sections
 and $\rho^{pp}$ ($\rho^{\bar{p}p}$) data 
in the interval $13 \le \sqrt{s} \le 62.5$ 
( $ 546 \le \sqrt{s} \le 1800$ GeV) }
\begin{tabular}{|l|l|l|l|}
$\sqrt{s}$ (TeV) &  $ C \ (GeV^{-2})$ & $\alpha^{-2} \ (GeV^{-2})$ & $\lambda $\\ \hline
          &   &      &      \\ \hline
13.8   &   9.97   &  2.092        & -0.094   \\ \hline
19.4   &  10.05   &  2.128        &  0.024   \\ \hline
23.5   &  10.25   &  2.174        &  0.025   \\ \hline
30.7   &  10.37   &  2.222        &  0.053   \\ \hline
44.7   &  10.89   &  2.299        &  0.079   \\ \hline
52.8   &  11.15   &  2.370        &  0.099   \\ \hline
62.5   &  11.50   &  2.439        &  0.121   \\ \hline
546    &  ???     &  ???          &  ???     \\ \hline
630    &  ???     &  ???          &  ???     \\ \hline
1800   &  ???     &  ???          &  ???     \\ \hline
   &  &  &   \\
\end{tabular}
\end{table}
\end{center}
\begin{center}
\begin{table}
\caption{Predicted $\sigma_{tot}^{pp}$ from fitting accelerator data:
 (a) extrapolation including data at $546$ GeV and $1.8$ TeV (the two first values are 
interpolations).
 (b) extrapolation with data at $\sqrt s \leq 62.5$ GeV;
 Experimental values are displayed in Table 1}
\begin{tabular}{|l|l|l|l|r|}
\multicolumn{1}{c}{$\sqrt s$ (TeV)} & \multicolumn{2}{c}{$\sigma_{tot}^{pp}$ (mb)} & 
\multicolumn{2}{c}{$\sigma_{tot}^{pp}$ (mb)} \\ \hline
          &   &      &   &    \\ \hline
0.55 & Intrp. &  $61.91_{-1.1}^{+1.2}$    & Extrp. &  $69.39_{-7.4}^{+8.4}$       \\ \hline
1.8   & Intrp. &  $76.78 \pm 1.4$        & Extrp. &  $91.74_{-14.7}^{+16.9}$    \\ \hline
14    & Extrp. & $110.49_{-3.1}^{+3.2}$    & Extrp. &  $143.86_{-33.5}^{+38.6}$   \\ \hline
30   & Extrp. & $125.63_{-4.1}^{+4.3}$    & Extrp. &  $167.64_{-42.6}^{+48.9}$    \\ \hline
40   & Extrp. & $131.71_{-4.6}^{+4.8}$    & Extrp. &  $177.23_{-46.3}^{+53.1}$    \\ \hline
100   & Extrp. & $152.45_{-6.2}^{+6.4}$     & Extrp. &  $210.06_{-59.1}^{+67.6}$   \\ \hline
\multicolumn{3}{c}{(a)}               & \multicolumn{2}{c}{(b)}  \\ 
   &  &   &   &   \\
\end{tabular}
\end{table}
\end{center}
\begin{center}
\begin{figure}[b!]
\centerline{\epsfig{file=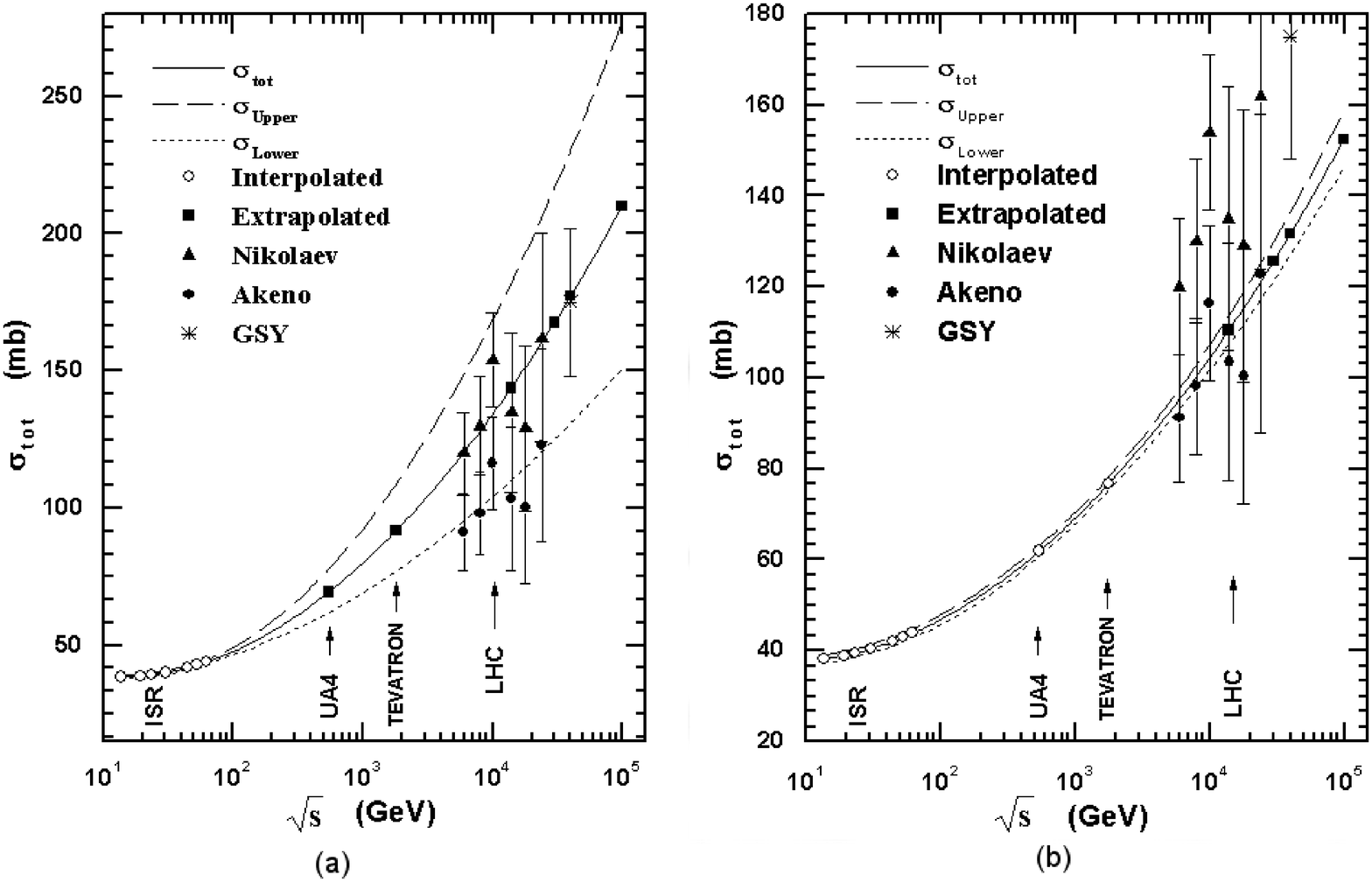,height=110mm,width=153mm}}
\caption{Predictions (black squares) of $\sigma_{tot}^{pp}$
(a) width data at $\sqrt s \leq  62.5$ GeV;
(b) including data at 546 GeV and 1.8 TeV. Open circles denote the interpolations.
 Notice the different vertical scales.}
\end{figure}
\end{center}
\section{Conclusions:}
It has been shown in this work that highly confident predictions of high energy 
$\sigma_{tot}^{pp}$ values are strongly dependent on the energy range 
covered by experimental data 
and the available number of those data values. 
In particularly, we show that if we 
limit our study of determining $\sigma_{tot}^{pp}$ at cosmic ray energies 
from extrapolation of 
accelerator data of $\sqrt s \leq  62.5$ GeV, then results are compatible 
with most of cosmic ray 
experiments and other prediction models, because the predicted error band 
is so wide that covers 
their corresponding error bands (Fig. 2a). However, 
as x
the included data in our calculations 
extends to higher energies, that is, when all experimental  
available data is taken into 
account, the estimated values for $\sigma_{tot}^{pp}$ obtained 
from extrapolation  and those 
obtained from cosmic ray experiments are only  compatible, within the error 
bars, with the Akeno results (Fig. 2b). It should be noted 
that our predictions are compatibles 
with other prediction studies \cite{Augier1}.   
Taken all these convergences at face 
value, as indicating the most probable $\sigma_{tot}^{pp}$ value, 
we conclude that if predictions 
from accelerator data are correct, hence, it should be 
of great help to normalize the 
corresponding values from cosmic ray experiments, as for instance 
by keeping the ($k$) parameter 
as a free one, as it is done for instance in  \cite{Block2}.
The $k$ value found will greatly help the tunning 
of the complicated Monte Carlo calculations used to evaluate the development of the showers
induced by cosmic rays  in the upper atmosphere. 
In summary, extrapolations from accelerator data should be used to constraint cosmic ray 
estimations.


\begin{references}
\bibitem{Engel1} 
Engel R., Gaisser T.K., Lipari P., Stanev T., {\it Phys. Rev. D} {\bf 58}, 014019 (1998).
\bibitem{hires} 
See http://sunshine.chpc.utah.edu/research/cosmic/hires/ 
\bibitem{Auger} 
The Pierre Auger Project Design Report. {\it Fermilab report} (Feb. 1997).
\bibitem{DL1} 
Donnachie, A and Landshoff, P.V. {\it Phys. Lett. B} {\bf 296}, 227 (1992). 
\bibitem{Augier1} 
Augier, C. et al  {\it Phys. Lett. B} {\bf 315}, 503 (1993a). 
\bibitem{Block1}
M.M.Block, F. Halzen and T. Stanev, Phys. Rev. Lett. {\bf 83}, 4926 (1999).
\bibitem{G1} 
Glauber, R.J., 1956, {\it Lectures in Theoretical Physics} (Reading: Interscience, N.Y.). 
\bibitem{GM1} 
Glauber R.J., Matthiae G., {\it Nucl. Phys. B} {\bf 21}, 135 (1970). 
\bibitem{GV1} 
Glauber, R.J., Velasco, J. {\it Phys. Lett. B} {\bf 147}, 380 (1984). 
\bibitem{MM1} 
Martini, A.F. and Menon, M.J. {\it Phys. Rev. D} {\bf 56}, 4338 (1997).
\bibitem{Amaldi1}
Amaldi, U. et al {\it Phys. Lett. B} {\bf 44}, 11 (1973). 
\bibitem{ISR1} 
Amendolia, S.R. et al {\it Phys. Lett. B} {\bf 44}, 119 (1973). 
\bibitem{Amaldi2} 
Amaldi, U. et al. {\it Phys. Lett.  B} {\bf 66}, 390 (1977).
\bibitem{UA41} 
Bozzo, M. et al {\it Phys. Lett. B} {\bf 147}, 392 (1984). 
\bibitem{E710} 
Amos, N. et al, {\it Phys. Rev. Lett.} {\bf 63}, 2784 (1989).
\bibitem{CDF} 
Abe et al. {\it Phys. Rev. D} {\bf 50}, 5550 (1994).
\bibitem{GM2} 
Matthiae, G. {\it Rep. Prog. Phys.} {\bf 57}, 743 (1994). 
\bibitem{Blois99} 
{\it Proc. IXth Blois Workshop}, {\bf 1999}, on Elastic and Diffractive Scattering, Protvino, 
Russia. 
\bibitem{FM1} 
Froissart, M. {\it Phys. Rev.} {\bf 123}, 1053 (1961); Martin, A. 1966, Nuovo Cimento 42, 930 
(1966). 
\bibitem{Pryke} 
C. L. Pryke,  e-Print Archive: astro-ph/0003442
\bibitem{Hillas} 
A. M. Hillas, Nucl. Phys. B (Proc. Supp.) {\bf 52B}, 29 (1997).
\bibitem{Fletcher} 
R. S. Fletcher et al., Phys. Rev. D {\bf 50}, 5710 (1994).
\bibitem{Akeno1} 
Honda, M.  et al {\it Phys. Rev. Lett.} {\bf 70}, 525 (1993).
\bibitem{FE1} 
Baltrusatis, R.M. et al {\it Phys. Rev. Lett.} {\bf 52}, 1380 (1984).
\bibitem{FE2} 
Baltrusatis, R.M. et al {\it Porc. 19Th ICRC}, La Jolla (1985).
\bibitem{Niko} 
Nikolaev, N.N. {\it Phys. Rev. D} {\bf 48}, R1904 (1993).
\bibitem{GSY} 
Gaisser, T.K., Sukhatme, U.P., Yodh, G.B. {\it Phys. Rev. D} {\bf 36}, 1350 (1987).
\bibitem{Augier2} 
Augier, C. et al {\it Phys. Lett.  B} {\bf 316}, 448 (1993b).
\bibitem{PP1} 
Perez-Peraza, J., Gallegos-Cruz, A., Velasco, J., Sanchez-Hertz, A. Faus-Golfe, A. and 
Alvarez-Madrigal, M. {\it ``Prediction of p-p total cross-sections with highly confident 
uncertainty band", in AIP Proceedingg of the Metepec, Puebla}, International Workshop on 
Observing ultra high energy cosmic rays from space and earth, august (2000).  
\bibitem{PP2}
Velasco, J., Perez-Peraza, J., Gallegos-Cruz, A., Alvarez-Madrigal, M., Faus-Golfe, A. and 
Sanchez-Hertz, A. {\it Proc. 26$^{\ th}$ ICRC, UTAH}, {\bf 1}, 198 (1999).
\bibitem{Block2}
M.M.Block, F. Halzen and T. Stanev, Phys. Rev. D {\bf 62}, 077501 (2000).
\end{references}
\end{document}